# Machine Learning for Semi-Automated Meteorite Recovery

Seamus Anderson[1*], Martin Towner[1], Phil Bland[1], Christopher Haikings[2,3], William Volante[4], Eleanor Sansom[1], Hadrien Devillepoix[1], Patrick Shober[1], Benjamin Hartig[1], Martin Cupak[1], Trent Jansen-Sturgeon[1], Robert Howie[1], Gretchen Benedix[1], Geoff Deacon[5]

[1]Space Science and Technology Center, Curtin University, GPO Box U1987, Perth, WA 6845, Australia
[2]Spectre UAV Concepts, 191 St Georges Terrace, Perth, WA 6000, Australia
[3]Amotus Pty Ltd, Level 25/71 Eagle St, Brisbane City, QLD 4000, Australia
[4]Department of Psychology, Clemson University, 418 Brackett Hall, Clemson, SC, 29634
[5]Western Australian Museum, 49 Kew St, Welshpool, WA 6106, Australia
[*]Corresponding author: E-mail: seamus.anderson@postgrad.curtin.edu.au.

## Abstract

We present a novel methodology for recovering meteorite falls observed and constrained by fireball networks, using drones and machine learning algorithms. This approach uses images of the local terrain for a given fall site to train an artificial neural network, designed to detect meteorite candidates. We have field tested our methodology to show a meteorite detection rate between 75-97%, while also providing an efficient mechanism to eliminate false-positives. Our tests at a number of locations within Western Australia also showcase the ability for this training scheme to generalize a model to learn localized terrain features. Our model-training approach was also able to correctly identify 3 meteorites in their native fall sites, that were found using traditional searching techniques. Our methodology will be used to recover meteorite falls in a wide range of locations within globe-spanning fireball networks.

## Introduction

Fireballs and meteors have been observed since antiquity by Chinese, Korean, Babylonian and Roman astronomers (Bjorkman 1973), while meteorites and their unique metallurgical properties have also been known and used by various cultures around the world from Inuit tools (Rickard 1941) to an Egyptian ceremonial dagger (Comelli et al. 2016 ), their connection to each other and to asteroids as source bodies was not proposed until the 19th century, with the fall of the l'Aigle meteorite (Biot 1803; Gounelle, 2006). Since this link was established, meteorites have, and continue to offer unique insights into the history of the solar system, as well as the contemporary characteristics, both physical and chemical, of asteroids, the Moon and Mars. Unfortunately, the overwhelming majority of these ~60,000 samples have no spatial context since their falls were not observed, leaving their prior orbits uncharacterized. Less than 0.1 % of meteorites in the global collection were observed well enough during their atmospheric entry to properly constrain their orbits (Meier 2017; Borovička et al. 2015; Jenniskens et al. 2020). This ultra-rare subset of meteorites afford some of the most valuable information pertaining to extra-terrestrial geology, since their physical and geochemical properties, along with their orbital histories can be combined to characterize the nature of asteroid families, and therefore possible parent bodies, that inhabit the same orbital space.

The best methodology for recovering meteorites with corresponding orbits, utilizes fireball camera networks, which use automated all sky camera stations in an overlapping arrangement such that a potential fireball can be imaged by two or more stations. From these observations, scientists can triangulate an atmospheric trajectory, from which a pre-entry orbit and a fall area can also be calculated. The first success of such a system was demonstrated in Czechoslovakia in 1959, with the

Pribram meteorite fall (Ceplecha 1961). This event spurred the establishment of the Czech fireball network (Spurný et al. 2006), along with multiple networks across the globe (McCorsky and Boeschenstein, 1965; Halliday et al. 1996; Oberst et al. 1998; Bland 2004, Brown et al. 2010; Colas et al. 2014; Devillepoix et al. 2020).

### The Desert Fireball Network

An ideal location for one of these networks was determined to be the Nullarbor region in Western and South Australia, due to its low humidity, sparse vegetation and typically clear skies (Bevan and Binns, 1989), thus the Desert Fireball Network (DFN) was born (Bland et al. 2012). Since its inception, the DFN has been responsible for the recovery of four confirmed meteorite falls: Bunburra Rockhole, Mason Gully, Murrili, and Dingle Dell (Spurný et al. 2012, Towner et al. 2011, Bland et al. 2016, Devillepoix et al. 2018), all of which have well constrained orbits. To date, the network covers approximately 30% of the land mass of Australia, with more than 50 camera stations (Howie et al. 2017). On average it observes 300 fireballs per year, typically 5 of which result in a meteorite fall.

For every fireball event observed by multiple camera stations, the bright flight trajectory is triangulated. If a terminal mass (meteorite fall) is predicted, we incorporate wind models into Monte Carlo simulations in order to estimate the likely fall area (Sansom et al. 2015, Howie et al. 2017, Jansen-Sturgeon et al. 2019). Since the fireball appears only as a streak of light, crucial attributes pertaining to the object such as size, mass and shape, are all co-dependent variables. This means that the predicted fall location results in a line, along which all of these parameters vary (Sansom et al. 2019). Inherent uncertainties and gaps in reported wind conditions at altitudes all along the flight, lead to a variation of ~250 m on either side of this fall line. Each predicted fall zone is entirely dependent on the conditions of the fireball, though typical events can result in a fall zone 2-4 $km^2$ in area. The decision to search for a particular meteorite is dependent on many factors, from the geometry and confidence of the trajectory triangulation, to local terrain features and geographic accessibility. Once the team has determined the fidelity of the triangulation and conditions of the fall area itself, a searching trip is commissioned to look for the fallen meteorite.

### Meteorite Recovery

Traditional methods for meteorite recovery include two main strategies, petal searching and line searching. Petal searching involves sending individuals out from a central point, walking alone or in small groups in a loop, typically a few km long, looking for and collecting meteorites along the way. This method generally covers a larger area but comes with a higher risk of missing meteorites in the area covered. Since this method is usually implemented in strewn fields or in areas with older surface ages and higher meteorite density, such as the Nullarbor (Bevan 2006), where the objective is to recover older meteorite finds, missing some meteorites is less detrimental.

Alternatively, line searching is more useful when trying to recover a meteorite fall with a well constrained fall line, like those observed through a fireball network. The DFN implements this strategy by assembling searchers in a line, spaced 5-10 m apart, then sweeping the area ~250 m on either side of the fall line on foot. This approach is usually able to cover 1-2 $km^2$ for each trip, assuming 6 people search 8 hours per day, for 10 days. The Antarctic Search for Meteorites (ANSMET) uses a similar method, only they are not restricted by a fall line, and instead cover the area with greater spacing while mounted on snowmobiles (Eppler 2011). The benefit of the line method is higher fidelity on the area covered, due to overlapping fields of view by the searchers, although generally, less area is covered with this method.

When considering both the number of meteorites found by the DFN, and the number of searching trips it has commissioned (4 and ~20 respectively), the success rate remains at ~20%. This relatively low rate combined with the cost (~20,000 AUD) of sending six people on trips for two weeks at a time necessitates an improvement in the meteorite recovery rate, particularly due to the establishment and expansion of the Global Fireball Observatory (Devillepoix et al. 2019).

**Previous Drone-Meteorite Recovery Methodologies**

The gargantuan strides that have been made in the last 10 years in the manufacture of high resolution DSLR cameras and commercial drones capable of carrying them, have opened the possibility of using both to aid in the recovery of meteorites. Previous attempts have been met with mixed to promising results. Moorhouse (2014) in his honors thesis, explored the possibility of using a hyper-spectral camera mounted to a drone to look for the possibly unique spectral signature of meteorites. This approach is unfeasible in our framework since the best hyper-spectral cameras are prohibitively expensive (>100,000 AUD), and more importantly, would limit our area coverage rate to little more than 0.1 km$^2$ per day, due to low spatial resolution in the camera. Further complications arise from the fact that Moorhouse used a spectral library of meteorite interiors, rather than meteoritic fusion crust, which is what would appear on the surface of fresh meteorite falls. Although meteorite fusion crusts could have a unique spectra compared to typical terrestrial environments, this is not explored in his work. Su (2017) focused on the feasibility of using magnetic sensors suspended from a drone, but this method would preclude us from finding non-magnetic meteorites, and also limited our area coverage to less than 0.1 km$^2$ per day. This approach would also be the most susceptible to obstacles on the ground and changes in local elevation, since they prescribe flying at a 2 m altitude.

Citron et al. (2017) relied on an RGB camera to survey an area, and used a machine learning algorithm to identify likely meteorites in the images. Their tests resulted in a meteorite detection rate of 50% and encountered a false positive rate of ~4 per 100 m$^2$. These results are very promising and seem to be limited mainly by the performance of the drone and camera hardware. The other limitation is false positives, and more importantly, how to separate them from from promising meteorite candidates. This is a crucial detail when considering that a typical fall line (>2 km$^2$), analyzed with their model, could have over 100,000 detections, all of which must be examined by a human in one way or another.

The work of Zender et al. (2018) also employed an RGB camera to image meteorites in native backgrounds. They showed the unique reflectance signature of meteorites in each color channel and created an algorithm to detect these signatures. This approach was able to detect half of their test-meteorites, though it did suffer from a high rate of false positives.

AlOwais et al. (2019) also used an RGB camera, while additionally investigating the utility of a thermal imaging system. They also train a number of neural networks to detect meteorites within images. One of their chief priorities was to create such an image processing system that would fit on-board their surveying drone. With this in mind, they elected to use transfer learning (Pan and Yang 2009) from a handful of smaller pre-trained neural networks, to detect meteorites. Their training resulted in a high model accuracy using images taken from the internet, as well as photo-shopping cropped meteorite-images onto terrain backgrounds. These results are promising and await validation in the form of field tests.

Our previous work on drone-meteorite recovery is described in Anderson et al. (2019). In this previous iteration, we trained a machine learning model on a synthetic dataset. We created it by taking survey images from a drone, splitting them into tiles, then overlaying the tiles with cropped meteorite images. Although training on these tiles yielded a high training accuracy, it was unable to consistently identify real test-meteorites placed on the ground, mostly likely because the training data lacked the native lighting conditions and shadows seen in the real test-meteorite images.

Here, we report on updated methods to achieve a practical system for recovering meteorites using drones and machine learning. Such a system must fulfill the following 6 requirements to be effective:

1) Survey at least 1 km$^2$/day,
2) Meteorite recovery chance (success rate) greater than 50%
3) Portable to different terrains/locations
4) Deployable by 3 people or less (1 vehicle)
5) Total cost <40,000 AUD (2 traditional searching trips)
6) Data processing rate equal to data surveying rate (including model prediction, and false-positive sorting)

**Methods**

**Drone and Camera Hardware**

In recent years, the number of options for consumer and commercial drones has grown dramatically, with many options including fixed-wing, multi-copter, vertical takeoff/landing, and even blimps. The designs with the most flight-proven heritage, at our price range, are fixed-wing and multi-copters. Our previous experience has shown that fixed wing models produce too much image blur and are unable to achieve a meaningful image resolution due to lower limits on most models' cruising altitude. Given these constraints we chose a DJI M600 drone to perform full scale tests as well as surveys of our fall sites. This drone was able to carry our camera and gimbal payload with mass to spare for possible later upgrades. It was also able to perform pre-planned survey flights, with meter-scale GPS precision, for more than 15 min at a time.

We also decided to use an RGB camera, since these systems are both scalable and widespread, as opposed to thermal or hyper-spectral cameras that are more expensive, specialized and are only capable of smaller spatial resolutions. We specifically chose a Sony A7R Mk. 3 (42 MPixel), with a 35 mm Lens, set to take images with a 1/4000 sec exposure, at f/4.5, and an ISO of 320. The total cost of the camera, drone, batteries, and accessories was 30,000 AUD, well below the 40,000 AUD limit we self-imposed in Criteria (5).

We used the DJI GO 4 app to control the drone manually during training data collection flights, while the survey flights were planned and executed using the DJI GS Pro app. With this equipment, we conducted tests at varying altitudes and determined that an image resolution of 1.8 mm/pixel (15 m altitude) would be sufficient to detect most of our typical meteorite falls (0.3 kg – 4 kg). This would allow meteorites to appear in the image between a size of 18 and 60 pixels in diameter. Using this fixed resolution value, we found that this system could survey approximately 1.3 km$^2$/day, when we flew nearly continuously for 7 hrs per day, easily fulfilling Criteria (1). Although we had 12 hours of daylight at the time of our full scale test, we found that surveying less than 2.5 hours after sunrise or before sunset produced long shadows that resulted in an unacceptably high rate of false positives.

**Machine Learning Software**

Since a meteorite would appear to be small (18-60 pixels) relative to the total size of the image (42 MPixel), we decided to split each image into 125x125 pixel-tiles with a stride of 70. This allowed a meteorite to fully appear in at least one tile, to maximize the chance of detection and minimize false positives. These tiles were then fed to a binary image classifier, a type of deep convolutional neural network, to separate uninteresting terrain (0) from meteorite suspects (1). We considered any prediction over 0.9 confidence to be a detection, or a possible meteorite.

We implemented our neural network by constructing a model in python using tensorflow (Abadi et al. 2015) and keras (Chollet 2015), the architecture of which is shown in Table 1. Although a sufficiently deep architecture is important when training a neural network, the training data itself is the most important factor, especially in our case where we trained the model from randomized initial weights (from scratch), rather than using a pre-trained network. This means that for a given fall site we needed numerous, diverse examples of both True (meteorite) and False (non-meteorite) tiles. The False tiles were relatively easy to assemble. We simply took a survey of an area without any meteorites, and made all of the images into False tiles.

The True tiles required a bit more effort. Since all the meteorites we would be searching for would have fallen within the last 10 years, they would all have intact, dark fusion crusts covering their surface. Fresh meteorites such as these also tend to be minimally altered, making them more analytically valuable to the meteorite community. This consideration limited the number of real meteorites that were available to us to use in data sets. To artificially bolster the number of True tiles we could generate, we also used stones with desert varnish surfaces, a dark, slightly shiny exterior that develops on some rocks in hot deserts (Engel and Sharp 1958), as 'synthetic' meteorites. At each site we also found stones that had a plausible meteoritic shape (non-jagged and without a noticeable elongated axis) and painted them black. Using this combination of fusion-crusted meteorites, desert varnish stones and painted stones, we always had enough samples to make a substantial number of True tiles.

Table 1. Meteorite-Detecting Neural Network Architecture

| Layer type | Filters (conv.) / neurons (dense) | Kernel size | Stride size | Activation function |
|---|---|---|---|---|
| **Convolutional 2D** | 30 | 3 | 1 | Rectified Linear Unit |
| **Batch Normalization** | | | | |
| **Max Pooling** | | 2 | 2 | |
| **Convolutional 2D** | 60 | 3 | 1 | Rectified Linear Unit |
| **Batch Normalization** | | | | |
| **Max Pooling** | | 2 | 2 | |
| **Convolutional 2D** | 120 | 2 | 2 | Rectified Linear Unit |
| **Batch Normalization** | | | | |
| **Max Pooling** | | 2 | 1 | |
| **Convolutional 2D** | 240 | 3 | 1 | Rectified Linear Unit |
| **Batch Normalization** | | | | |
| **Max Pooling** | | 2 | 1 | |
| **Flatten** | | | | |
| **Dense** | 100 | | | Rectified Linear Unit |
| **Dropout** | 30 % | | | |
| **Dense** | 50 | | | Rectified Linear Unit |
| **Dense** | 25 | | | Rectified Linear Unit |
| **Dense** | 5 | | | Rectified Linear Unit |

| Dropout | 30% | |
| Dense | 5 | Rectified Linear Unit |
| Dense | 1 | Sigmoid |

Our procedure for making these tiles is illustrated in Figure 1. Step 1 consisted of laying out the stones in a line at the fall site, spaced more than 1 m apart, and then imaging them with the drone. This line could be either in the fall zone, or just outside of it, in order to train the model on similar backgounds. We gave a 1 m separation to ensure that two stones would not appear in the same tile, when we augmented the data later on. We found the best way to accomplish this stone-imaging was for one person to walk ~3 m parallel to the line of black stones, and point to each one, while another person manually flew the drone at the prescribed survey altitude, following the first person. Physically pointing out each individual stone allowed us to annotate each stone only once, avoiding a possible double appearance of a particular stone in both the training and validation sets. For Step 2, we drew a tight bounding box around each stone and recorded the box's height, width and position in the image. These annotations were completed using ImageJ (Schneider et al. 2012). We typically laid out ~100 stones at a time; 15% of these stones and their resultant tiles are set aside for validation, not used in training. This ensured that the validation set only consisted of stones that the model had never seen, as opposed to unseen permutations of stones that the model was already familiar with.

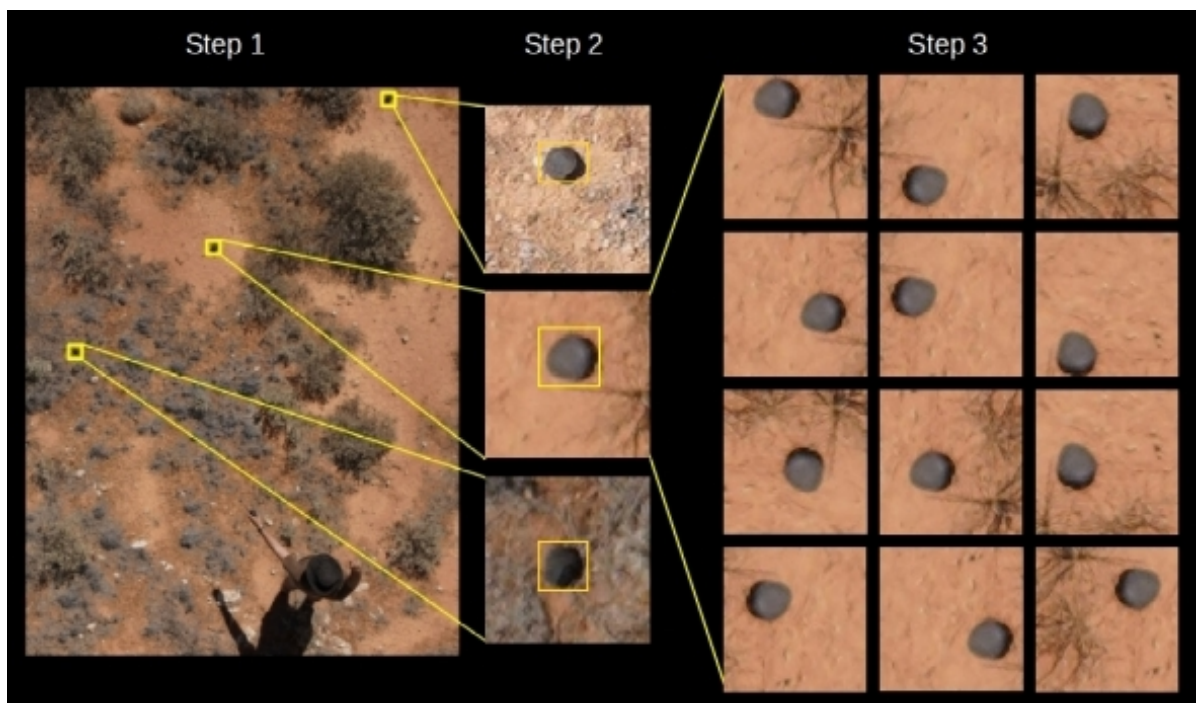

Fig. 1. Our workflow for obtaining meteorite training data. Step 1 consists of laying out the stones on the ground >1 m apart, and imaging them at a 1.8 mm/pixel resolution. Step 2 shows how we record the position, height and width of each rock in the full-sized image, by drawing a bounding box. Step 3 is where we generate the tiles to be used for training and validation.

At Step 3, we took each annotation, in both the training and validation sets, and strode by 15 pixels in both axes over each meteorite, creating a new tile at each stride, while keeping the stone fully in the tile frame. Each of those tiles was then rotated in intervals of 90 degrees and saved for each permutation. These strides and rotations force each rock to appear in nearly every position of a tile, without any preference in local directionality, i.e. shadows and windblown vegetation. We repeat this data-collection process at different times of the day, at different sections of the fall line, to include as much variety as possible. Details like these are crucial when making a widely generalized training set.

This process ideally generates ~50,000 True tiles for the training set. To assemble the False tiles, we flew the drone 350 m, parallel to the fall line, taking images all along the way. By splitting the images into tiles, we generated ~2,500,000 False samples. The process of laying out stones, imaging everything, and making the annotations typically took an hour.

Since dramatically unbalanced datasets can negatively affect training (Miroslav and Matwin 1997), we could only train with as many False tiles as we had True tiles, to keep a 1:1 ratio. A simplified example of unbalanced datasets is a training set containing 1 True and 99 False samples. Mathematically speaking, the shortest path to the model achieving a high accuracy would be for it to label everything false, resulting in an accuracy of 99%. Obviously, this kind of solution is useless, which is why we must maintain a ratio as close to 1:1 as possible. We did this by randomly selecting 50,000 False tiles from the pool of 2.5 million, and combining them with the 50,000 True tiles, to form the whole training set. We also included ~8,000 False tiles from the 2.5 million into the validation set, ensuring they did not also appear in the training set.

We trained on our dataset for 150 epochs (rounds of training), using a batch size of 250, with 400 steps per epoch, which ensured that each tile is seen by the model once per epoch. The validation set was evaluated at the end of each epoch, also using a batch size of 250 with 64 steps. For smaller datasets, we adjusted our batch size and steps per epoch such that the product of these two values equaled the size of the training set.

Once the model completed training, we judged its utility based on its meteorite detection chance, and rate of false positives. The meteorite detection chance was determined by predicting on each of the True tiles in the validation set, and dividing the number predictions over 0.9 confidence by the total number of tiles. This provided a metric for how well the model could correctly identify new black rocks that it had never seen. For the false positive rate, we wanted to obtain a more widespread and representative value that would reflect model performance across the whole fall zone. So we randomly selected 50 images from the survey of the fall line and predicted on them with the model, recording the average number of detections across all the images.

**Model Detection Sorting Interface**

An issue we anticipated with any model we would train, was the processing of false positives. Even in best case scenarios, where we assume a model accuracy of 99.999%, with ~8,500 tiles per image and ~650 images per flight, a model would return approximately 5,500 detections per flight, and more than 150,000 per fall line. Thus, we required a tool to help searchers efficiently examine each of these model-detections, and determine which of these were obvious false positives and which ones required further investigation. We created a graphical user interface in python using the Tkinter module to accomplish this task (Figure 2).

The program displays nine detections at a time, in a 3x3 grid pattern. Each grid space is mapped to numbers 1-9 on a standard keyboard's keypad (1 for lower left, 5 for middle-center, 9 for upper right). Each detection is displayed such that the frame is centered on the detection tile, outlined in a yellow box (~25 cm on one side), and extends 70 pixels beyond the target tile, to give the user context of the larger area. Below the grid, 3 images of meteorites are displayed, scaled from the smallest to the largest meteorite possible for that fall site (lowest mass with iron density, to highest mass with chondritic density, respectively). This allows users to easily reference how big a meteorite should appear in the tiles. If the user decides that the tile likely contains a meteorite, they press the number on the keypad corresponding to that grid space, before advancing to the next set. The program also allows the user to remove their responses from the current set of 9 tiles, as well as go back to the previous set.

   Through testing trails, we determined that the average user could sort through ~120 tiles per minute. Assuming 150,000 detections per fall line, the task of sorting through this data would take over 20 labor-hours. This problem of staying focused over long periods of time is known as 'vigilance ' by human factors psychologists, who have observed decrements in user performance over extended task sessions (See et al. 1995). To mitigate such decrements in vigilance, we ensured that each user would only sort for 20-minute increments. This was chosen as a conservative time limit according to Teichner (1974), who found the vigilance decrement to be fully observed 20-35 minutes into a task. Additionally, to reduce the consequence of individual errors, each tile was inspected by two separate users. We also anticipated that the overwhelming majority of detections would be false positives, thereby counter-productively enticing the users to speed through the tiles, without properly inspecting each one. The resulting consequence of such task parameters has been shown in signal detection literature to result in the missed detection of such rare signals (Stanislaw & Todorov 1999), in our case the user-detection of a meteorite. To combat this, we added a test function to the program, whereby each set of nine tiles had a uniform probability of containing 0, 1 or 2 test tiles, taken from the training set. This forced the user to slow down and select, on average, one tile per set, thus reducing the rarity of a "hit".

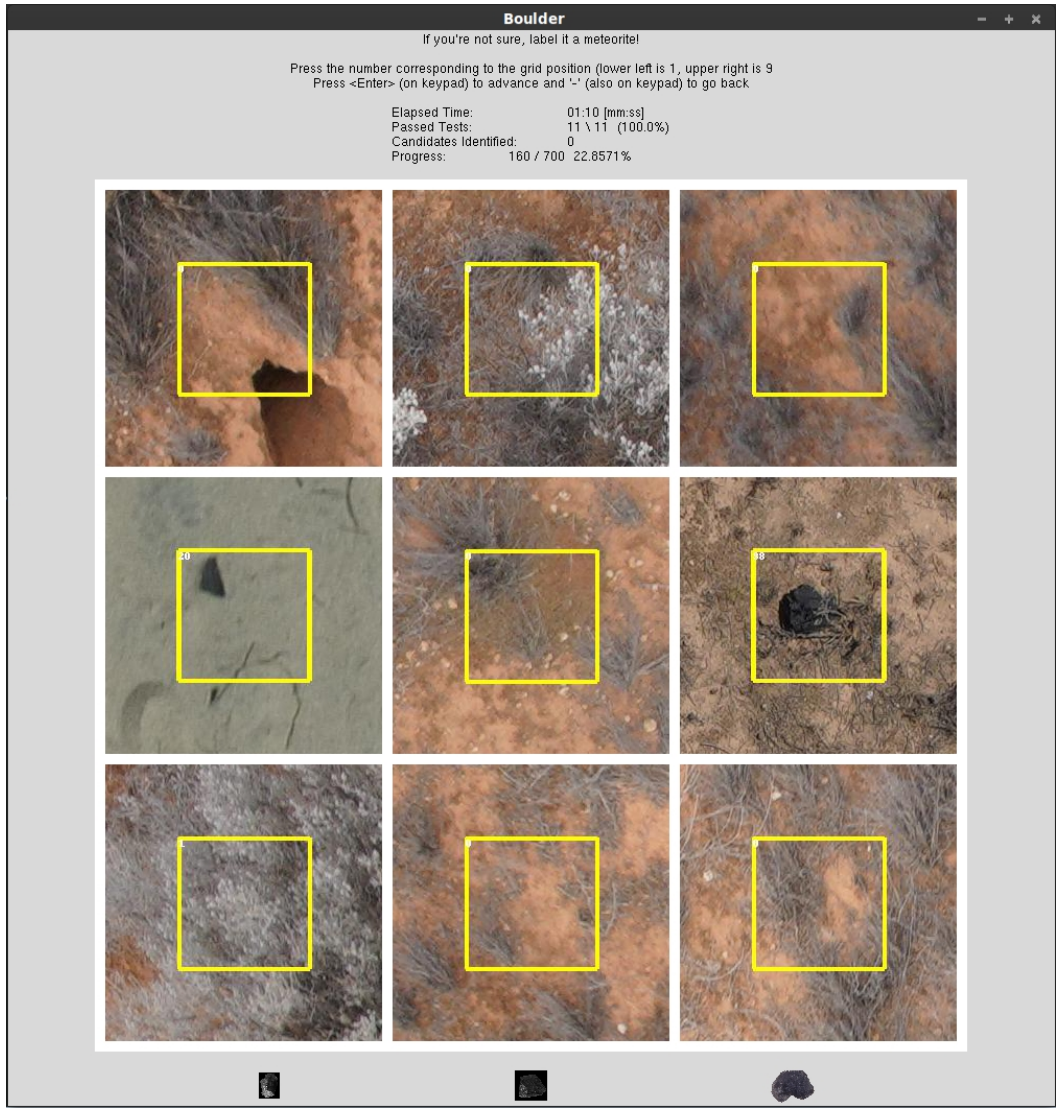

Figure 2. Our sorting user interface we designed to aid in the separation of candidates from false positives. Test tiles (True tiles from the model training set) for this set appear in the center-left and center-right positions. Users press the corresponding number on the keypad to mark the tile(s) as a likely meteorite.

A final failsafe was included in this sorting task, such that once the user missed two test tiles during a sorting session, the program would shut down, forcing the user to take a break. The user's score of successfully completed tests, along with the number of meteorite candidates identified, are shown at the top of the display. Both of these strategies; increasing the "hit" rate and providing performance feedback, have been shown to combat the vigilance decrement (Hancock et al. 2016).

Once two users sorted through the detections for a flight, we overlaid the original images with bounding boxes around meteorite candidates. We also set aside the false positives, so that we may use them for retraining if needed.

## Results

We conducted small scale tests of our methodology by visiting 4 sites in Western Australia and training a model at each location. Although they were not at real meteorite fall sites, they were all located within the DFN's operational area, and could conceivably be representative of future fall sites. These sites and the results from the models we trained for them are listed in Table 2. For these smaller tests, we only obtained training data for ~30 synthetic and real meteorites, and surveyed less than 0.1 km$^2$ at each site.

Table 2. Distinct models at various locations within the DFN. Model performance is dependent on the size of the training dataset.

| Location | (Lat, Long) | Total Number of Training Tiles | Training Accuracy | Meteorite Detection | False Positives (per image) |
|---|---|---|---|---|---|
| Ledge Point | (-31.151, 115.395) | 16,352 | 92.58 % | 68.5 % | 21.7 |
| Dalgaranga | (-27.635, 117.289) | 30,874 | 97.03 % | 85.6 % | 6.5 |
| Lake Kondinin | (-32.496, 118.192) | 32,348 | 96.85 % | 86.7% | 5.1 |
| Balladonia | (-32.370, 124.790) | 98,470 | 98.73 % | 93.2 % | 1.3 |

We also conducted a full test of our methodology by visiting one of our fall sites, DN150413_01, North-East of Forrest Airport, Western Australia (30.764 S, 128.184 E). We obtained training data for this fall site at different times during the day (morning, cloudy mid-morning, midday, early afternoon, and late afternoon). Over the course of two days we also surveyed 2 km$^2$ of the fall zone, so that we could identify meteorite candidates for secondary inspection in an upcoming expedition. We also placed 4 painted rocks, unseen by the model, within the survey area and recorded their GPS coordinates. This served as a test of our ability to use the model to correctly identify a meteorite candidate, correctly sort it using the user interface, and accurately correlate the image's GPS coordinates to the those recorded by our handheld unit. During the survey, each of three team-members were assigned a distinct role during survey-flight operations. The first team-member's job was to fly the drone and calibrate the camera, the second oversaw data collection and backups on the computer, and the third was responsible for cooling and charging the batteries.

When we returned from the field, we trained a model on our RTX 2080 Ti (11 GB RAM) GPU, with an Intel i9-9000 CPU for approximately 3 hours (150 epochs). This resulted in a final training accuracy of 99.07% and a validation accuracy of 98.65%. Furthermore, we achieved a meteorite detection chance of 98.71%, and a false positive rate of 2.5 per image. Using the trained model, the detection algorithm was able to process 1 day's images in 22 hours. The model returned a combined total of 92,595 detections for the two-day survey, which we were able to sort through in 12 hours, excluding breaks.

Sorting through all of our detections yielded 752 meteorite candidates, some of which are shown in Figure 3. Of the four test rocks we laid out, we successfully located three of them (by comparing GPS coordinates) using our prescribed searching methods, meaning that we successfully met fulfilled criteria: (2).

Four months after this initial trip, when the COVID-19 travel restrictions were lifted in Western Australia, we revisited the same site North-East of Forrest Airport. We began by inspecting ~20 of the 749 candidates in-person and noticed that they generally belonged to one of two populations: dark stones (most likely iron-rich siliceous rock), and small holes in the ground (<7 cm in diameter) most likely made by small animals. The small hole population was far more numerous than the dark stone group and were easy to distinguish in the images, once we knew which features to look for. We then sorted through the remaining ~700 candidate images and narrowed the list to 32 candidates that did not appear to be holes in the ground. Unfortunately after inspecting these remaining candidates, we found that none of them were meteorites.

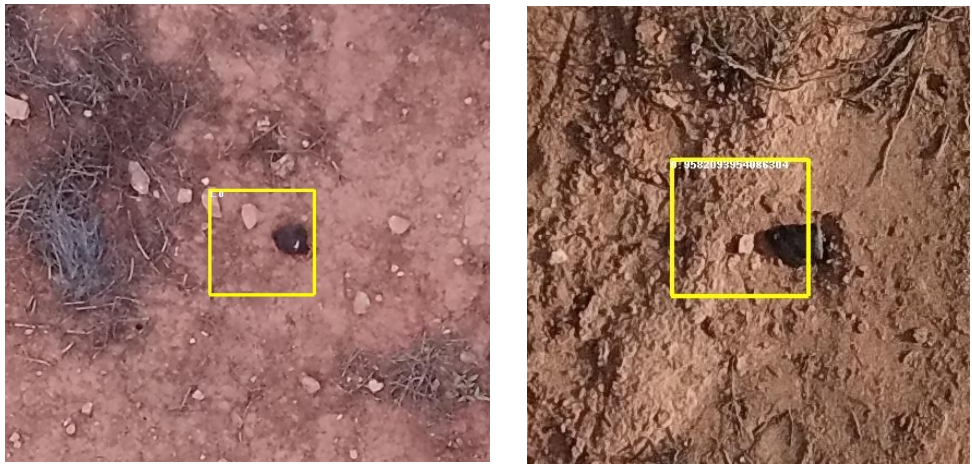

Fig 3. The meteorites recovered North-West of Forrest Airport (left) and South of Madura (Right). Although these two were not recovered using the full surveying methodology, they serve as valuable demonstrations for the feasibility of our approach.

On this same follow-up trip, we also visited a second, separate fall site located North-West of Forrest Airport. For this site, we employed the traditional line searching technique and found the meteorite on the afternoon of the first day. Using our Mavic Pro drone, we took ~100 images of this meteorite (Fig. 3, left) from a top down view, with heights ranging from 1 to 30 m. We also generated training data at this site, trained a model, and used it to predict on 86 of these images (those in which the meteorite was between 10 and 80 pixels in diameter). The model was able to correctly identify the meteorite in 84 of the 86 images, or 97%.

During a separate trip, whereby two members of our research group were scouting a third fall site South of Madura, Western Australia, for an upcoming six-person searching trip, they discovered the meteorite in question (Fig. 3, right), on the dirt road which roughly bisected the predicted fall line. They also used the Mavic Pro to take images of the meteorite from altitudes of 2 to 30 m, and created training data on-site. When they returned from the trip, we trained a model and predicted on the 27 meteorite images, finding that the model correctly identified the meteorite in 24 of the images (88% success rate). At the writing of this manuscript, these two meteorites have not yet been registered with the Meteoritical Bulletin, as their classifications are forthcoming.

An additional and final test of our approach involved using our Forrest-NE model to predict on a drone-image of an older meteorite find, shared with us by a volunteer meteorite hunter who regularly searches in the Nullarbor. We found that the model correctly identified the old chondrite with a prediction value of 1.0: a perfect match.

**Discussion and Future Work**

Our smaller tests (Table 2) show that more training data makes for a more robust model in terms of both meteorite detection and false positives, reinforcing the notion that more training data makes for a better model. These tests also showcase the portability of our methodology, accounting for variations in available training data, which successfully satisfies Criteria (3).

The results of the full test, while not a total success, are a promising prospect for the future of meteorite recovery. Not only is this methodology capable of locating test meteorites analogues, it is able to cover a fall zone nearly 6 times faster than a traditional line-search, when accounting for invested labor. If we assume that in the future we would predict on images and sorting through detections in the field, this rate of data processing can keep pace with data collection through a combination of switching sorting users and simply taking breaks, satisfying the final outstanding Criteria (6).

There are two possibilities as to why the full test did not result in a complete success of recovering the meteorite. The first explanation is that our methodology failed at some stage of the searching, whether the model failed to detect the meteorite, or we failed to label the detection as a candidate. The second possibility was that we did not cover enough of the fall line. Since the initial surveying trip was limited to two days, we were only able to cover 2 km$^2$ of the entire 5 km$^2$ fall zone. Our other successes with the models correctly identifying two fresh falls and one old find, all in situ, lead us to believe that the second explanation is more likely. For this reason, we plan on returning to the Forrest-NE fall site and surveying the remainder of the fall zone.

We will also embark on an extensive surveying campaign of all of our meteorite fall sites. We initially plan on training a new model for each fall site, using randomly initialized weights. Though as we gain more training data from a range of diverse fall sites, we will investigate the possibility of combining data sets and training a 'base model' who's final weights will then be used as the initial weights for each new model we train. This future approach may improve the generalizability of our models and reduce training time on-site.

The python software that we have created, as well as our trained model weights, will be made available to collaborators upon request, so that the entire meteoritics community can benefit from this new method of semi-automated meteorite recovery.

*Acknowledgements- We would like to thank Robert Tower for providing us with additional drone-survey images. This work was funded by the Australian Research Council (Grant: DP200102073).*